%% file: main.tex
\documentclass[acmsmall,screen,nonacm]{acmart}

\usepackage{tikz}
\usepackage{amsmath}
\usepackage{shortcut}
\usepackage{filecontents}
\usepackage{url}

\setcopyright{none}
\acmDOI{}
\acmISBN{}
\acmYear{2027}
\acmConference[FSE 2027]{ACM International Conference on the Foundations of Software Engineering}{July 12--16, 2027}{Shenzhen, China}

\begin{document}

\emergencystretch=1em

\author{Ziyi Liang}
\authornote{Both authors contributed equally to this work.}
\affiliation{
  \institution{Huazhong University of Science and Technology}
  \country{China}
}
\email{d202180584@hust.edu.cn}

\author{Zichuan Li}
\authornotemark[1]
\affiliation{
  \institution{University of Illinois Urbana-Champaign}
  \country{USA}
}
\email{zichuan7@illinois.edu}

\author{Bin Yuan}
\affiliation{
  \institution{Huazhong University of Science and Technology}
  \country{China}
}
\email{yuanbin@hust.edu.cn}

\author{Deqing Zou}
\affiliation{
  \institution{Huazhong University of Science and Technology}
  \country{China}
}
\email{deqingzou@hust.edu.cn}

\author{Luyi Xing}
\affiliation{
  \institution{University of Illinois Urbana-Champaign}
  \country{USA}
}
\email{lxing2@illinois.edu}

\title{Mission (Im)Possible: Towards Autonomous Discovery of Autopilot Logic Flaws in Open-Source Drone Frameworks}

\begin{abstract}
\input{sections/00.abstract}
\end{abstract}

\maketitle

\section{Introduction}
\label{sec:intro}
\input{sections/01.intro}

\section{Background}
\label{sec:background}
\input{sections/02.background}

\section{Motivating Example}
\label{sec:motivation}
\input{sections/03.motivation}

\section{\textsc{I}mpossible \textsc{M}ission \textsc{F}inder (\tool)}
\label{sec:design}
\input{sections/04.design}

\section{Implementation}
\label{sec:implementation}
\input{sections/05.implementation}

\section{Evaluation}
\label{sec:evaluation}
\input{sections/06.evaluation}

\section{Case Study}
\label{sec:case_study}

\input{sections/07.case_study}

\section{Discussion}
\label{sec:discussion}
\input{sections/08.discussion}

\section{Related works}
\label{sec:related_works}
\input{sections/09.related_works}

\input{sections/11.appendix}

\section{Conclusion}
\label{sec:conclusion}
\input{sections/conclusion}

\bibliographystyle{IEEEtran}

\bibliography{ref}

\end{document}

%% file: sections/00.abstract.tex
Autonomous drones are increasingly deployed in safety-critical tasks, yet their autopilot software can still contain subtle logic flaws that lead to unsafe behaviors. 
These flaws are difficult to discover with existing techniques because they often depend on semantic relationships across modules, commands, and runtime state, and may not manifest as crashes or other easily detectable failures.
With recent advances in large language models (LLMs), we observe that, when equipped with essential domain knowledge and carefully crafted context, LLMs demonstrate strong capabilities in understanding the high-level semantics of autopilot systems, making them suitable for logic flaw discovery. 
However, LLMs are limited by their context and can be biased to the vulnerabilities frequently occurs in training dataset, and logic flaws may not always lead to crashes and other observable behaviors, which means it require higher-level semantic bug oracles to identify its existence.
To bridge the gap, we present \textsc{\textbf{I}mpossible \textbf{M}ission \textbf{F}inder} (\tool), a multi-agent system combining static analysis, dynamic execution and log analysis, that supports fully automated \textit{identify}, \textit{test}, and \textit{validate} logic flaws in autonomous vehicles. 
\tool primarily contains three LLM agents, it first uses the auditor agent to propose logic flaw hypothesis based on carefully-crafted program slicing, then it launches tester agents to write and execute test cases for each candidate, and finally it spawns the log analyzer agent to validate the execution log against the flaw hypothesis.
This enables \tool dynamically validate bug oracles in a case-by-case manner, making it suitable for discovering logic flaws that do not have a unified pattern.
We evaluated \tool on two widely used open-source autopilot systems, \px and \ardupilot, leading to
105 previously unknown logic flaws, that can cause unsafe behaviors including mission interruption failures, invalid actions, entering into restricted fly zones etc.

%% file: sections/01.intro.tex
Autonomous drones have been increasingly deployed into various safety-critical scenarios~\cite{DJI-agriculture,AIRIAL-PX4,HarrisAerial-PX4,Germanrail-airialuas}, making their security increasingly important. Security flaws and vulnerabilities in drones can lead to operational failures and safety-critical issues,
such as navigation errors or collisions with other objects~\cite{drone-collision}.
While existing security research has demonstrated that both
open-source~\cite{opensource-vulnerable1,opensource-vulnerable2} and commercial
drones~\cite{schiller2023DJIdroneID} are vulnerable to various attacks~\cite{command-inject, mem-corrupt, authentication}, the logic issues, that can also lead to serious real-world impacts, have been largely overlooked.
Logic flaws are difficult to find since they often emerge from cross-module state propagation, implicit design assumptions and runtime semantics. 
For example, vulnerabilities in the mission execution module can result in race conditions, where improper synchronization of shared state variables causes the drone to ignore or mishandle critical commands. In \PX, a lack of synchronization for the \texttt{vehicle\_status} variable between the \texttt{Commander} module and the \texttt{Navigator} can cause the drone to occasionally fail to pause its mission when receiving a PAUSE command (see \S~\ref{sec:case_study}).

Recent LLMs have demonstrated strong code reasoning capabilities and have been widely adopted in both software development~\cite{claude-code,codex} and security analysis~\cite{red-anthropic,vim-claude,kim:atlantis}. Although they are effective at detecting vulnerabilities with well-defined patterns, applying them to discover logic flaws in autopilot frameworks remains challenging due to several reasons.
First, discovering logic flaws requires high-level semantic understanding across multiple modules, which demands carefully constructed contextual information for LLMs, which is crucial for open-source autopilot frameworks due to their massive codebases (more than 6M lines of code for both \ardupilot and \px).
Second, LLMs are structurally vulnerable to hallucinations and can be biased by their training data~\cite{cloak}. Directly applying them to logic flaw discovery may therefore produce an overwhelming number of false positive results.
Third, unlike memory corruption or taint-style vulnerabilities, logic flaws do not follow unified patterns and often do not result in crashes or other easily observable behaviors, making it difficult to design generic bug oracles.
To address these challenges, we present \textsc{\textbf{I}mpossible \textbf{M}ission \textbf{F}inder} (\tool), a multi-agent system that can automatically \textit{identify}, \textit{test}, and \textit{validate} logic flaws in open-source autopilot frameworks.
We equip LLM agents with tools for querying carefully constructed program slices, enabling them to reason about semantic context and generate high-quality findings. The agents then automatically synthesize test cases to execute and validate auditing hypotheses. Finally, we introduce a log analysis agent that examines runtime execution logs and determines whether the observed behaviors are consistent with the hypothesized logic flaws.

We evaluated \tool on two widely used open-source autopilot frameworks, \px and \ardupilot, and
confirmed 105 previously unknown logic flaws after manual verification.

\vspace{3pt} \noindent \textbf{Contributions.} We summarize our contributions as follows:

\noindent $\bullet$ \textit{New method.} We propose a novel LLM-powered approach for discovering logic bugs in drone autopilot systems that cause unexpected behaviors. Our approach leverages LLMs to bridge the gap between high-level code logic semantics and runtime behaviors. Our key insight is that modern LLMs can learn implicit logic requirements as long as provided with carefully crafted context, making them suitable for recognizing unexpected behaviors.

\noindent $\bullet$ \textit{New tool.} 
We design and implement \tool, a fully automated multi-agent system to discover logic flaws in open-source autopilot frameworks, which combines static code analysis with LLM-powered test generation and execution with a LLM-empowered log analysis. \tool provides practical automation for developers, handling everything from test case generation to log analysis without manual intervention.
We will release all the implementation and evaluation scripts used in \tool.

\noindent $\bullet$ \textit{New results.}
We evaluate the effectiveness of \tool on two widely adopted autopilot frameworks, \px and
\ardupilot. Our evaluation shed light on 105 previously unknown logic bugs, 
showing the effectiveness of \tool.

%% file: sections/02.background.tex
\subsection{MAVLink Protocol and MAV\_CMD}
\label{sec:bg_mavlink}

MAVLink~\cite{MAVLINK} is a lightweight messaging protocol that connects a
drone to external systems such as ground control stations, companion
computers, and developer tools. Major open-source autopilots, including
\px~\cite{px4-github} and \ardupilot~\cite{Ardupilot-github}, expose MAVLink
as their common operator-facing interface, which makes it a compatible
abstraction layer across different ground control station clients. 

The \mavlink protocol is structured around a packet-based design, where each message has a payload for specific commands or telemetry data and a checksum for ensuring data integrity. 
Internally, \mavlink facilitates cross-module data sharing, enabling seamless communication between different components such as the navigator, commander, and sensors.
And externally, \mavlink messages enable users to send commands and instructions to the drone, such as setting up missions or adjusting flight parameters. Furthermore, users can utilize \mavlink to set critical data in the drone, including home positions, GeoFence boundaries, and mission configurations, ensuring precise control and operational safety.

\subsection{Flying Modes and Missions}
\label{sec:bg_modes_missions}

A drone autopilot operates in one of several discrete \textit{modes}, each
determining how the vehicle responds to operator input and sensor data. Common
modes include: \textit{takeoff, hold or loiter, mission or auto, and return-to-launch (RTL)}.
The modes transit either when an operator issues
an explicit MAVLink command or when a safety condition triggers an automatic
switch, such as a fallback to RTL on low battery or loss of communication.

A \textit{mission} is a structured flight task that the autopilot executes
when placed in mission mode. Operators define a mission as an ordered
sequence of \textit{waypoints}, each specifying a geographic target by
latitude, longitude, and optionally altitude, and may interleave the
waypoints with actions such as commanding takeoff or landing. The mission
is uploaded over MAVLink as a sequence of mission-item messages; the
autopilot stores the items and, while in mission mode, executes them in
order.

\subsection{Indirect Data Flow in Autopilot Frameworks}
\label{sec:bg_indirect}

Within an autopilot framework, much of the data flow between modules is
mediated by an indirected layer rather than by direct function calls. \px
relies on the \textit{uORB} publish-subscribe system: modules publish typed
messages to named topics, and other modules subscribe to those topics. Because
topic names are strings resolved at runtime, the connection from a publisher
to a subscriber is not visible as a direct call in the source code.
\ardupilot uses different mechanisms, including a parameter store and
command-handler tables, but the same pattern of one external input fanning out
across multiple modules before reaching actuator outputs applies.

To make this fan-out concrete, consider a typical command such as
\texttt{MAV\_CMD\_DO\_REPOSITION}. After the dispatch step described in
\S~\ref{sec:bg_mavlink}, the command-handler publishes the parsed command
into a uORB topic such as \texttt{vehicle\_command}. The navigator subscribes
to \texttt{vehicle\_command}, processes the entry, and publishes a derived
setpoint topic such as \texttt{position\_setpoint\_triplet}; position-control
modules in turn subscribe to the setpoint topic and produce actuator outputs.
\ardupilot exhibits an analogous fan-out through parameter writes and
handler-table dispatch.

%% file: sections/03.motivation.tex
In this section, we use a real logic flaw in \px to motivate why autopilot mission logic requires semantics-aware, slice-guided analysis. Although the example comes from \px, it reflects a broader difficulty in reasoning about mission-state inconsistencies in drone autopilots.

\vspace{3pt}\noindent\textbf{Overview}.
The example shown in Figure~\ref{fig:motivating_example} originates from the \navi module of \px, which contains a vulnerability:
\textit{
If the \textbf{home\_position.altitude} is updated during flight, the target waypoint is not recalculated, but the reachability check uses the updated altitude. This state inconsistency can prevent the system from recognizing that the waypoint has been reached.
}

\input{code/motivating_example}

\vspace{3pt}\noindent\textbf{Vulnerability details.}
Specifically, when a user updates the home position via the ground control station, a
\texttt{MAV\_CMD\_DO\_SET\_HOME} command is issued. Upon receiving the command, the drone updates its internal
\texttt{\_home\_pos} topic with the new value. Within the \navi module, this value is retrieved through topic getter functions
(\one).
This updated home position further includes two distinct code slicings in the figure (\two\ and \three). In \two, home\_position
is used to calculate and publish the position\_setpoint\_triplet topic. In \three, it is
used to compute the distance between the drone's current location (retrieved via get\_global\_position \four) and
the target mission position. This distance is then stored in the\gbox{\texttt{dist\_z}} variable.
If this distance falls within a predefined threshold (\five), the corresponding waypoint flag variable is marked as reached (\six).
However, the publication of the mission target (\two) occurs only when users directly update the
mission or when the current mission waypoint is reached, shown in \seven.

Since the slicing \two\ (\texttt{set\_mission\_items} \eight\ inside the if block) is \textbf{not}
triggered immediately upon updating \texttt{\_home\_pos}, the low-level motor modules continue to
use previously published setpoints, i.e., stale coordinates.
In contrast, the slicing \three\ is invoked every time \texttt{Navigator::\allowbreak on\_\allowbreak active} is called
(\seven), so \texttt{mission\_item\_altitude} uses the updated \texttt{\_home\_pos}.
This discrepancy introduces inconsistency: the drone computes distances with an updated target
position but navigates according to outdated targets. As a result, \texttt{dist\_z} fails to satisfy
the acceptance condition (\five), and the system is unable to mark the waypoint as reached (\six).
Consequently, the drone continues to perform heading corrections toward an already-passed waypoint,
ultimately causing it to loiter indefinitely around invalid target coordinates.

\vspace{3pt}\noindent\textbf{Can LLMs help?}
To test whether current LLMs can identify mission-logic flaws in 
Figure~\ref{fig:motivating_example}, we ran a small preliminary study under two context settings:
(a) a manually crafted slice covering the relevant dataflow and control conditions, and (b) the full
navigator and mission source files, including both \texttt{.cpp} and \texttt{.h} files. We evaluated
four models (\texttt{GPT-5.4}, \texttt{GPT-5.4-mini}, \texttt{GPT-OSS}, \texttt{llama-4-scout})
each with three trials. With the prepared slice, \texttt{GPT-5.4} identified the issue in all
trials, while smaller models sometimes identified partial symptoms; with full files, none of the
models found the target inconsistency and instead reported simpler non-semantic issues such as
integer overflow. These results showed that \textit{current LLMs are capable of identifying logic
flaws in autopilot systems as long as they are given focused, high-quality semantically clean
context.} This motivates our design of the \tool, a multi-agent pipeline that can \textit{identify},
\textit{test}, and \textit{validate} logic flaws in autopilot frameworks, detailed in
\S~\ref{sec:design}.

%% file: code/motivating_example.tex
\begin{figure}[h]
    \centering
    \begin{minipage}{0.95\linewidth}
    \centering

\begin{lstlisting}[style=cppscript]
class Navigator : public ModuleParams
{
  // Getter & Setter of topics
  home_position_s             *(*@\tikzmark{get-home-position-define-start}@*)get_home_position()(*@\tikzmark{get-home-position-define-end}@*);
}
\end{lstlisting}

\begin{lstlisting}[style=cppscript,firstnumber=6]
void Navigator::on_active()
{
 if(*@\tikzmark{is-mission-reached-start}@*)is_mission_item_reached_or_completed()(*@\tikzmark{is-mission-reached-end}@*)
    || mission_sub_updated || _mission_waypoints_changed){
    (*@\tikzmark{set-mission-start}@*)set_mission_items()(*@\tikzmark{set-mission-end}@*);
  }
}
\end{lstlisting}

\begin{lstlisting}[style=cppscript, firstnumber=13]
float MissionBlock::get_absolute_altitude_for_item(mission_item_s &mission_item)
{
    // ...
    return mission_item.altitude + _navigator->(*@\tikzmark{get-home-position-start}@*)get_home_position(*@\tikzmark{get-home-position-end}@*)()->alt;  
}
\end{lstlisting}

\begin{lstlisting}[style=cppscript, firstnumber=18]
bool MissionBlock::mission_item_to_position_setpoint( mission_item_s &item, position_setpoint_s *sp) {
  //...
  (*@\tikzmark{get-abs-alt-use-start}@*)sp->alt = get_absolute_altitude_for_item(item)(*@\tikzmark{get-abs-alt-use-end}@*);
}

void Misson::(*@\tikzmark{set-mission-def-start}@*)set_mission_items(*@\tikzmark{set-mission-def-end}@*)()
{
  mission_item_to_position_setpoint(_mission_item, 
  &(*@\tikzmark{pos-sp-triplet-current-start}@*)pos_sp_triplet->current(*@\tikzmark{pos-sp-triplet-current-end}@*));
  //...
  (*@\tikzmark{publish-position-setpoint-start}@*)publish_position_setpoint_triplet()(*@\tikzmark{publish-position-setpoint-end}@*);
}
\end{lstlisting}

\begin{lstlisting}[style=cppscript,firstnumber=29]
bool MissionBlock::(*@\tikzmark{is-mission-reached-def-start}@*)is_mission_item_reached_or_complete(*@\tikzmark{is-mission-reached-def-end}@*)d(){
  (*@\tikzmark{mission-amsl-def-start}@*)mission_item_altitude_amsl(*@\tikzmark{mission-amsl-def-end}@*) = (*@\tikzmark{get-abs-alt-in-mission-item-reached-start}@*)get_absolute_altitude_for_item(*@\tikzmark{get-abs-alt-in-mission-item-reached-end}@*)(_mission_item);    
  dist = get_distance_to_point_global_wgs84(
    ..., 
    (*@\tikzmark{mission-amsl-start}@*)mission_item_altitude_amsl(*@\tikzmark{mission-amsl-end}@*), 
    (*@\tikzmark{get-global-pos-start}@*)_navigator->get_global_position()->alt(*@\tikzmark{get-global-pos-end}@*), 
    ...,
    &(*@\tikzmark{distz-assign-start}@*)dist_z(*@\tikzmark{distz-assign-end}@*)
  );
  
  // Additional checks ...
  if ( (*@\tikzmark{distz-check-start}@*)... && dist_z <= _navigator->get_altitude_acceptance_radius()(*@\tikzmark{distz-check-end}@*)) {
    // ...
    (*@\tikzmark{way-point-true-start}@*)_waypoint_position_reached = true;(*@\tikzmark{way-point-true-end}@*)
  }
}
\end{lstlisting}

\begin{tikzpicture}[remember picture, overlay]
  \highlighcodebox[black,thick]{get-home-position-define}
  \highlighcodebox[red]{get-home-position}
  \highlighcodebox[cyan, thick]{set-mission}
  \highlighcodebox[cyan, thick]{set-mission-def}
  \highlighcodebox[blue]{get-abs-alt-use}
  \highlighcodebox[blue]{publish-position-setpoint}
  \highlighcodebox[olive, thick]{is-mission-reached}
  \highlighcodebox[olive, thick]{is-mission-reached-def}
  \highlighcodebox[blue]{pos-sp-triplet-current}
  \highlighcodebox[red]{get-abs-alt-in-mission-item-reached}
  \highlighcodebox[red]{mission-amsl-def}
  \highlighcodebox[blue]{get-global-pos}
  \highlighcodebox[green, thick]{distz-assign}
  \highlighcodebox[green, thick]{distz-check}
  \highlighcodebox[green, thick]{way-point-true}

  \draw[->, dashed, black, line width=1pt] (get-home-position-define-box) to[out=270, in=90] (get-home-position-box) node[] {};
  \draw[->, blue, line width=1pt] (get-home-position-box) to[out=270, in=90] (get-abs-alt-use-box) node[] {}; 
  \draw[->, blue, line width=1pt] (get-abs-alt-use-box) to[out=180, in=180] (pos-sp-triplet-current-box) node[] {}; 
  \draw[->, blue, line width=1pt] (pos-sp-triplet-current-box) to[out=270, in=120] (publish-position-setpoint-box) node[] {}; 
  \draw[->, red, line width=1pt] (get-home-position-box) to[out=270, in=90] (get-abs-alt-in-mission-item-reached-box) node[] {}; 
  \draw[dashed,->, blue, line width=1pt] (publish-position-setpoint-box) to[out=180, in=180] (get-global-pos-box) node[] {}; 
  \draw[->, blue, line width=1pt] (get-global-pos-box) to[out=-90, in=0] (distz-assign-box) node[] {}; 
  \draw[->, red, line width=1pt] (mission-amsl-def-box) to[out=180, in=180] (distz-assign-box) node[] {}; 

  \node[above right=2pt of get-home-position-define-box] {\Large \one};
  \node[below right=0pt and -15pt of get-home-position-box] {\Large \two};
  \node[below right=0pt and -45pt of get-home-position-box] {\Large \three};
  \node[right=2pt of get-global-pos-box]{\Large \four};
  \node[above=1pt of distz-check-box]{\Large \five};
  \node[below=2pt of way-point-true-box]{\Large \six};
  \node[below left=-20pt and 4pt of set-mission-box]{\Large \seven};
  \node[below=0pt of set-mission-box]{\Large \eight};

\end{tikzpicture}

\caption{A Motivating Example Vulnerability: in the figure, home position topic affects both position setpoint and mission reached check. The problem is that the slicing marked with blue lines is not always triggered when home position updated}
\label{fig:motivating_example}

\end{minipage}
\end{figure}
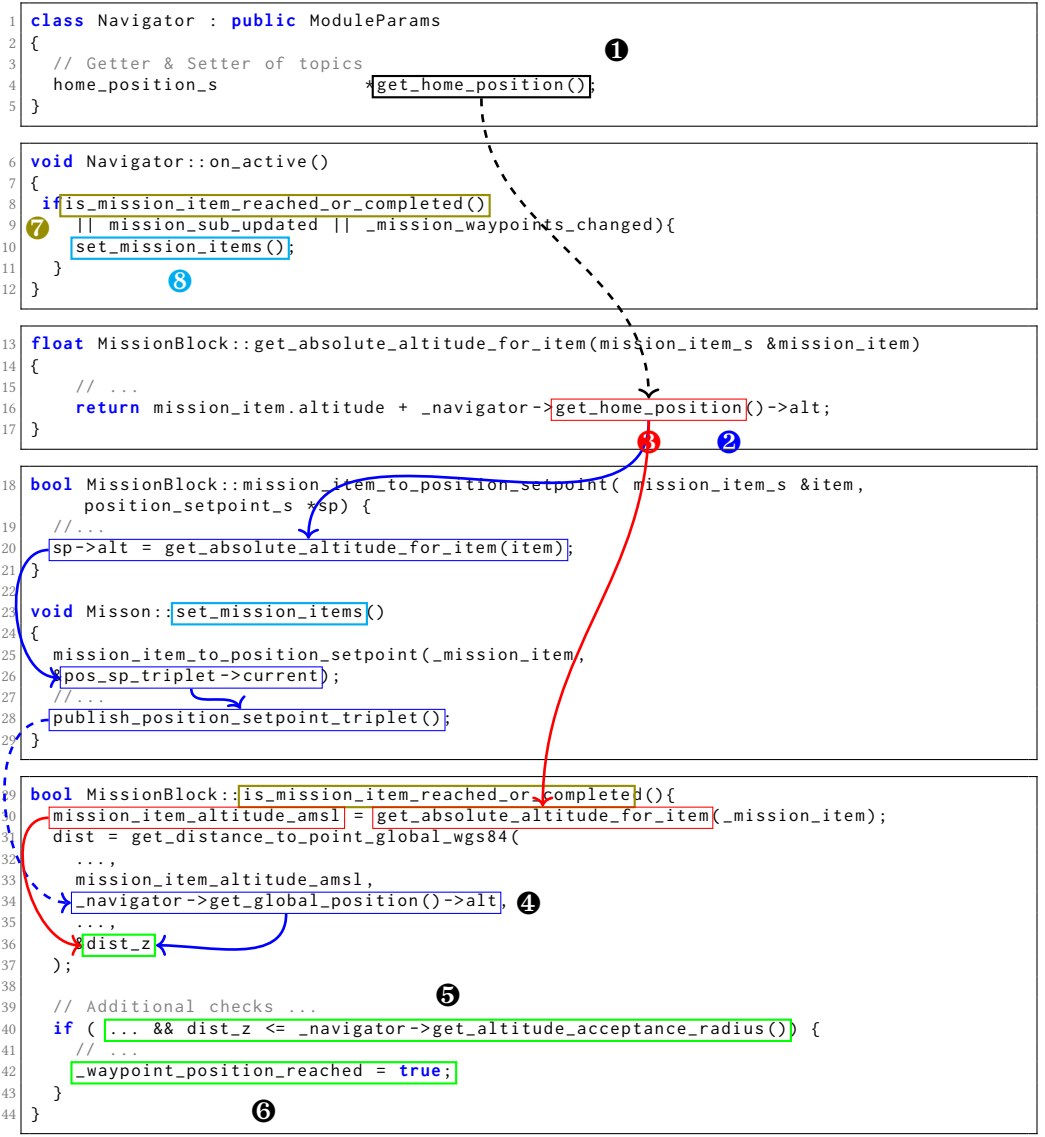

%% file: sections/04.design.tex
While current LLMs are powerful enough to reason and identify logic flaws in autopilot systems, it is still non-trivial to adopt them for logic bug discovery.
The agentic system needs to address the following three challenges:

\vspace{3pt}\noindent $\bullet$ \textbf{Challenge 1: How to prepare a good context for LLMs?}

\vspace{2pt}\noindent Recognizing the logic flaw in~\S~\ref{sec:motivation} requires the model to a)
trace and understand the dataflow of variables (home position and mission items), b) understand the
implicit relationships of setpoint and global position, and c) identify the difference of invocation
conditions. It's crucial to construct a high-quality context that contains necessary information a
model requires to identify the logic bug, while minimizing the token usage.

\noindent $\bullet$ \textbf{Challenge 2: How to minimize the model hallucination?}

\vspace{2pt}\noindent Even with good context, LLMs can still generate hallucinated results, making
it necessary for runtime verification. An automated agentic system therefore needs a way to drive
the autopilot through the candidate scenario and observe whether the hypothesis really can be
triggered.

\vspace{3pt}\noindent $\bullet$ \textbf{Challenge 3: How to detect the existence of logic bugs without explicit bug oracles?}

\vspace{2pt}\noindent
Different from traditional software bugs, logic flaws in autopilot systems are often semantic
inconsistencies that do not have a unified bug oracle. Detecting the existence of logic flaws
requires a dynamic bug oracle that can adapt to the candidate we want to verify.

\input{figures/architecture}

\vparstart{Our approach.}
To address these challenges, we design \textbf{I}mpossible \textbf{M}ission \textbf{F}inder (\tool), a multi-agent system powered by static analysis and large language models, which can automatically \underline{\textit{identify}}, \underline{\textit{test}}, and \underline{\textit{validate}} logic flaws in autopilot frameworks.
\tool primarily contains two phases: an offline phase that performs static analysis on autopilot
frameworks, constructs forward slicings for each MAVLINK command, and builds a \texttt{SliceDB}
(\S~\ref{sec:design-context}); and an online phase which contains three separate agents: an auditor
agent (\S~\ref{sec:design-context}) proposes candidate logic flaw hypothesis, a tester agent
(\S~\ref{sec:design-verify}) writes and executes simulator tests, and a log analyzer agent
(\S~\ref{sec:design-detect}) validates the execution logs to determine the existence of logic flaws.

\subsection{Preparing Analysis Context For LLMs}
\label{sec:design-context}

Identifying the logic flaw illustrated in \S~\ref{sec:motivation} requires giving the LLM both a focused cross function/module code context, and the necessary domain-specific semantic knowledge that connects components.
Without these inputs, LLMs can be biased by vulnerability patterns frequently occurred in its training dataset, such as integer overflow, buffer overflow, or missing input-validation checks, and rarely examine the higher-level semantic inconsistencies. 
To address this challenge, we design an auditor agent with static analysis tools specifically designed for autopilot frameworks. \tool a) performs the offline data-flow analysis and store code slicings into a database that the auditor agent can runtime interact with; and b) embeds domain-specific knowledge into auditor's prompt to make LLMs aware of the higher level semantic requirements and relationships between modules.

\vparstart{Forward Slicing Analysis.}
The precompute phase performs forward slicing on all MAVLink commands. Starting from a MAVLink command handler, \tool traces how that command's effect propagates through the rest of the autopilot framework.
Each analysis run begins at a MAVLink command's dispatch point in the MAVLink handler and follows the framework's data flow into the modules that consume the resulting state, producing a slice that consists of the relevant source functions, the control- and data-flow fragments within them, and the call chain that links them. The slice is the focused cross-function, cross-module context the audit agent needs to reason about a single command end-to-end.

A single MAVLink command typically has more than one dispatch point in the real codebase.
Listing~\ref{lst:dispatch_example} shows an example that \ardupilot matches
\texttt{MAV\_CMD\_DO\_SET\_HOME}: one inside \texttt{handle\_command\_int\_packet} that calls a
dedicated handler, and one inside the mission-item parser that inlines the body. 
\tool enumerates all such dispatch points for a command and forward-slices from each source, then
merges the slices into a single output slice. Listing~\ref{lst:slice_example} shows an example of
the slicing for \texttt{MAV\_CMD\_DO\_SET\_HOME} on \ardupilot.

\input{code/dispatch-tree-example}

\input{code/forward-slicing-example}

\vparstart{Indirect Dataflow in Autopilot Systems.}
In autopilot frameworks, \texttt{MAV\_CMD} packets can flow to consumers through indirect data flows. 
For instance, in the motivating example (\S~\ref{sec:motivation}), \texttt{get\_global\_position} returns a value that is gradually steered toward what an unrelated module previously published (\texttt{publish\_position\_setpoint\_triplet}), even though there is no dataflow edge directly connects the two functions. 
\tool primarily handles two types of indirect flow in \px and \ardupilot. 
The first type is uORB-style publish-subscribe communication, where one module can publish topic values
and other modules subscribe to those topics. When constructing the code slicing, \tool follows the
data flow of MAVLink command handlers, in which if stops at the publisher side,
\texttt{orb\_publish}, \tool would also slice the subscriber side from every static
\texttt{orb\_subscribe} on the same topic, and uses the topic name as the SliceDB join key.

Second, \tool also handles command aliasing across across modules.
In both frameworks, the same MAVLink command can be converted to different enum values, such as
a \texttt{MAV\_CMD\_*} can be converted to \texttt{NAV\_CMD\_*} in navigator module, and similarly
\texttt{VEHICLE\_CMD\_*} in other modules. 
For a single MAVlink command, \tool slices each of its alias command independently, stores the
slicings into the database, and merges them at retrieval time using a per-framework curated mapping
between equivalent command names.

\vparstart{SliceDB.}
The output of the forward slicing analysis is stored in a SQLite database, which we call SliceDB. During runtime, the auditor agent invoke tools to interact with the database.
This separation between an offline pre-compute phase and runtime semantic reasoning lets static analysis and the LLM each work on what they are good at, and save the dataflow analysis cost across audit runs. 

\vparstart{Agent Tools.}
The audit agent has three different kinds of tools:

\noindent $\bullet$ \textit{SliceDB Tools.} We provide the auditor agent two tools for interacting with SliceDB, a \texttt{query\_\allowbreak cmd\_\allowbreak slices} that takes a \texttt{MAV\_CMD} name as input and returns the forward code slicing stored in SliceDB; and a \texttt{query\_convergence} tool that takes two \texttt{MAV\_CMD} names as inputs and aims to address cases where an inconsistency only becomes visible across two commands meeting on a shared variable or state.
This tool internally checks the common sinks that both MAV\_CMD can affect and only returns the shared statements.

\noindent $\bullet$ \textit{Code Repository Exploration Tools.} 
Except for the slice-query tools, we also provide the audit agent the standard tools a coding agent typically uses: \texttt{read\_file}, \texttt{grep}, \texttt{glob}, and \texttt{bash} over the source code repository. 
The code slicing acts as a starting point, and we explicitly encourage LLMs to look into detailed functions and files. This makes the agent can dynamically explore and inspect more details about the siblings of the dataflow. For example, if it found a missing input validation, it could compare it with other sibling commands to see if there are similar issues.

\noindent $\bullet$ \textit{Sub-Agent Tools.}
Following a popular feature that state-of-the-art coding agents~\cite{claude-code, codex} support, we make auditor agent a coordinator, who can directly invoke other agents as subagents. This includes the tester agent (\S~\ref{sec:design-verify}) and the log analyzer (\S~\ref{sec:design-detect}), as well as auditor agent itself. That means, the auditor agent can invoke a subagent sharing the same structure and tools as itself, and delegate it for some subtasks. 
With these interfaces, we make auditor a fully autonomous pipeline that can keep iterating and dynamically invoke the testing and log analyzing process.

\subsection{Verifying with Runtime Execution}
\label{sec:design-verify}
The candidates produced by the audit agent are statically LLM-generated hypotheses, not confirmed bugs.
These hypotheses can subject to false positives and model hallucinations biased by frequent patterns in training dataset.
To address this challenge, we design a tester agent that converts candidate hypothesis into test cases, execute it inside simulators, and captures the execution log for further validation. 
To help the agent better write test cases, we a) design a testing harness package that exposes
unified API interfaces for both \px and \ardupilot, which lets the agent write test cases without
the need of understanding detailed framework APIs; and b) adopt a hypothesis dependency filter
to filter out cases cannot be validated inside the simulator and a semantic checks after each
attempts to ensure test cases can be validated and correctly reflect auditor's hypotheses.

\vparstart{Unified Testing Interface.}
The interface that the tester sees is modeled from the ground control station's perspective: it exposes the operations a human operator can perform over MAVLink protocols, such as uploading a mission, setting a No-Fly Zone, issuing a command, setting a parameter, etc. 
The same set of API interfaces can be used for both \px and \ardupilot, with the framework-specific requirements hidden inside the package details.
To make it easier for agent to use these APIs, we deliberately do not reuse the framework's own test surfaces, such as \px's MAVSDK and \ardupilot's \texttt{autotest} framework.
For example, MAVSDK is written in C++ and an agent needs to carefully initialize C++ objects with correct types and properly invoke APIs with expected types and objects, which can often cause errors. 
To upload a Mission in MAVSDK, the agent needs to properly initialized a set of structures and understand each parameter's meaning and their default values (shown in Listing~\ref{lst:mavsdk_example}). 
A single mis-filled field can silently invalidate the test or change its behavior.
Our wrapper, instead, uses Python and restricts the interfaces to basic types (e.g., list, str, int, etc). 
The underlying testing harness package handles the MAVLink struct construction and field initialization.
The agent only need to express high level semantics of API sequences. 
Listing~\ref{lst:upload_mission} shows our interfaces to upload a mission item, we carefully constructed docstrings and API documents that describes the API interface's design popurse, the meaning of each parameter, and their corresponding type hints.

When receiving a testcase, the testing harness package launches the simulator, crafts the MAVLink packets, and interacts with the simulator just like a ground control station.

\input{code/mavsdk_testcase}

\input{code/upload_mission}

\vparstart{Tester Agent.}
Given a candidate hypothesis from the audit agent, the tester writes a Python test using the unified interface. 
Before the test case is dispatched to the testing package, the tester internally runs a semantic alignment check on its attempts, which invokes LLM to check whether the testcase properly reflects the auditor's hypothesis and whether it could plausibly distinguish the bug from a normal run.
If the test is rejected, a rational would be returned to the tester for revision. 
Only after the test passes this check is it dispatched to execution, if it keeps failing, the testing will be aborted.

\subsection{Validating Existence of Logic Flaws}
\label{sec:design-detect}

Prior vulnerability discovery approaches based on fuzzing or static analysis normally depend on unified bug oracles, such as sanitizer exceptions, deterministic patterns, etc. so they detect only the limited classes of vulnerability that fit those oracles and cannot capture high-level semantic inconsistency.
To address this challenge, we use the LLM itself as a bug oracle and design a log analyzer agent inspired by Agent-Driver's use of ego-state information~\cite{agentdriver} in autonomous driving. Our analyzer maintains an ego state summarizing the vehicle's current status and processes the execution log entry by entry to update the state.
Given the execution log and the bug hypothesis, \tool a) first preprocesses the execution logs to only keep the necessary entries; b) collapses long silent duplicate log entries with a sliding window to keep LLM token cost manageable; and c) querying LLMs to detect whether current sliding window triggers the anomaly according to current ego state.

\vparstart{Log Preprocessing.}
Since \px and \ardupilot use different log formats, we first pre-process the logs to extract shared core elements, such as position and velocity components (\texttt{xyz}, \texttt{vx, vy, vz}).
In the preprocessing step, \tool also prompt LLMs to ask it to decide what entries should be kept and recorded as part of the ego-state.
A fixed ego-state can potentially be unable for analyzing the logic.
We let the agent to select the entries whose values are needed to judge the candidate's hypothesis, and those topics become the dynamic part of the ego state. 
This step is what lets \tool support dynamic discovery of new logic-flaw classes without rewriting the framework; the log preprocessing automatically specializes the ego state to whatever the audit agent currently produces.

\vparstart{Sliding Window over Silent Logs.}
A few-second SITL execution logs can contain thousands of entries. Since the autonomous vehicles typically record execution logs at a fixed sampling frequency, most of these entries are minor changes of the same state where most entries are identical and only a few types got changed.
Thus, feeding all the contents to LLMs can be costly and analyzes too much irrelevant logs as the bug hypothesis intend.
We therefore maintain a sliding window over the each tracked ego-state field. Only when the window has accumulated enough meaningful state changes, that is, when at least we have observed 10 entries of changes on ego-state fields, does the analyzer trigger an LLM call. 
This design significantly reduces 97\% of the overall LLM calls and help the log analyzer focus on high level semantics.

\vparstart{Ego State.}
In \tool, the ego state summarizes the vehicle's current status for execution-log analysis.
The key idea is to maintain a ego state of vehicle at a specific time point. Each time new logs coming, the agent invokes the LLM to interpret and parse the incoming logs and determine abnormal behaviors.
The loop terminates when a) either the inconsistency is observed, or b) all the contents in the log have been analyzed.

%% file: figures/architecture.tex
\begin{figure*}[t]
   \centering 
   \includegraphics[width=.95\linewidth]{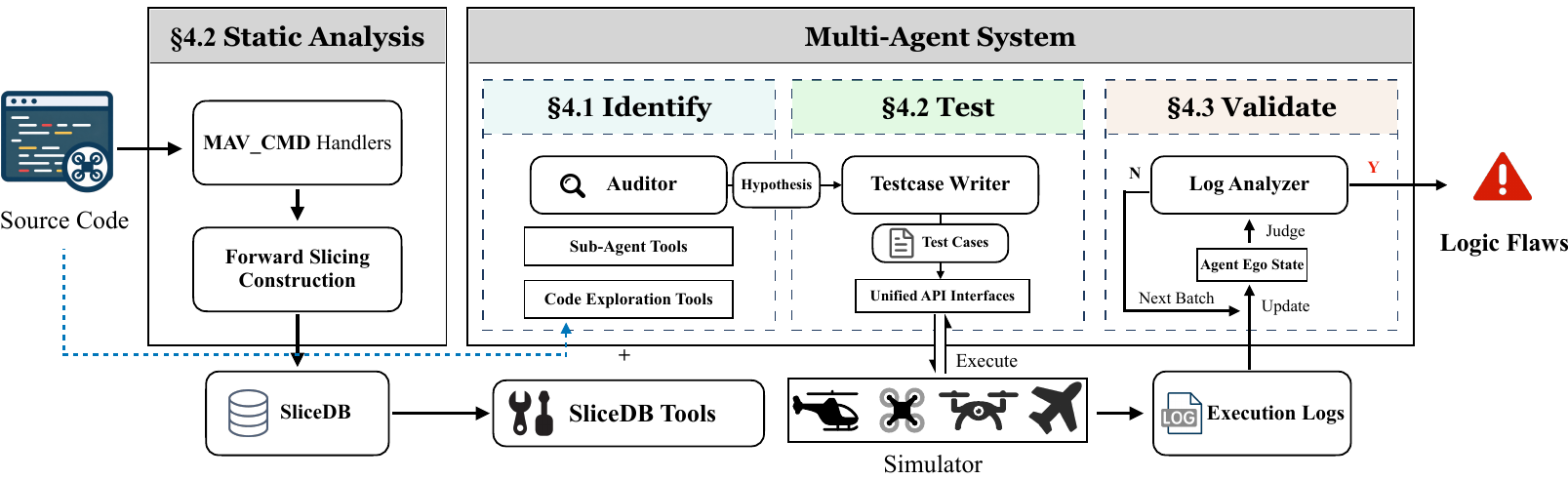}
   \caption{Overview of IMF. It first leverage static analysis to construct forward slicings of each \texttt{MAV\_CMD} handlers (\S~\ref{sec:design-context}), then the multi-agent system automatically proposes candidate logic flaws (\S~\ref{sec:design-detect}), writes and executes testcases in the simulators (\S~\ref{sec:design-verify}), and finally analyzes the execution logs to validate the existence of vulnerabilities (\S~\ref{sec:design-detect})}
   \label{fig:architecture}
\end{figure*}

%% file: code/dispatch-tree-example.tex
\begin{figure}[t]
\begin{lstlisting}[
  emph={MAV_CMD_DO_SET_HOME},
  emphstyle={\color{myblue}\bfseries},
  label={lst:dispatch_example},
  caption={Dispatch sites for \texttt{MAV\_CMD\_DO\_SET\_HOME} in \ardupilot.}]
MAV_CMD_DO_SET_HOME  (ArduPilot)

libraries/GCS_MAVLink/GCS_Common.cpp:5753
  // in GCS_MAVLINK::handle_command_int_packet
  case MAV_CMD_DO_SET_HOME:
      return handle_command_do_set_home(packet);   
      // -> handler at GCS_Common.cpp:5489

libraries/AP_Mission/AP_Mission.cpp:1232
      // in AP_Mission::mavlink_int_to_mission_cmd
      case MAV_CMD_DO_SET_HOME:
          cmd.p1 = packet.param1;
          break;
\end{lstlisting}
\end{figure}

%% file: code/forward-slicing-example.tex
\begin{figure}[t]
\begin{lstlisting}[basicstyle=\ttfamily\scriptsize, frame=single,
  label={lst:slice_example},
  caption={Example slice returned by SliceDB for \texttt{MAV\_CMD\_DO\_SET\_HOME}
  on \ardupilot's \texttt{COMMAND\_INT} ingress. The slice records the
  dispatch chain into the handler, the handler body, and the downstream
  functions reached on the forward taint trace. Other ingress paths and
  other framework targets produce additional, separately stored slices.}]
ardupilot/1: MAV_CMD_DO_SET_HOME
============================================================
Dispatch chain:
  GCS_MAVLINK::handle_command_int
    -> GCS_MAVLINK::handle_command_int_packet
      -> GCS_MAVLINK::handle_command_do_set_home

Handler: GCS_MAVLINK::handle_command_do_set_home
  libraries/GCS_MAVLink/GCS_Common.cpp:5489-5510
  MAV_RESULT handle_command_do_set_home(...) {
    // function body ...
  }

Reached functions on forward slice (taint-traced):
  is_zero 		libraries/AP_Math/ftype.h:63
  ModeAuto::do_set_home ArduCopter/mode_auto.cpp:2006
  ...
\end{lstlisting}
\end{figure}

%% file: code/mavsdk_testcase.tex
\begin{figure}[tb]
\centering
\begin{lstlisting}[basicstyle=\ttfamily\scriptsize, frame=single, style=hlstyle,
  label={lst:mavsdk_example},
  caption={Example of uploading a GeoFence using MAVSDK, the parameter of interfaces need to be properly initialized with correct types}]
TEST_CASE("Control Allocation - Remove one motor", "[controlallocation]")
{
    //...
|    AutopilotTester::MissionOptions mission_options;|
|    mission_options.rtl_at_end = false;|
|    mission_options.relative_altitude_m = flight_altitude;|

    AutopilotTesterFailure tester;
    tester.connect(connection_url);
    tester.wait_until_ready();
    // Configuration
    tester.set_param_ca_failure_mode(1);
|    tester.prepare_square_mission(mission_options);|
    //...
}
\end{lstlisting}
\end{figure}

%% file: code/upload_mission.tex
\begin{figure}[H]
\centering
\begin{lstlisting}[basicstyle=\ttfamily\scriptsize, frame=single,
  label={lst:upload_mission},
  caption={Uploading a custom mission via our harness: a single
  \texttt{upload\_mission} call carrying a \texttt{list[dict]} of
  basic-typed fields covers any MAV\_CMD --- no per-command struct
  setup, no test scaffolding.}]
//...
drone.upload_mission([
{"command": 16,  "lat": -35.36, "lon": 149.16, "alt": 20}, # NAV_WAYPOINT1
{"command": 16,  "lat": -35.41, "lon": 149.36, "alt": 20}, # NAV_WAYPOINT2
])
//...
\end{lstlisting}
\end{figure}

%% file: sections/05.implementation.tex
We implemented \tool as two separate Python packages. A multi-agent system package (\texttt{imf\_agent}, 11,952 LoC) built on top of LangGraph~\cite{langgraph}, and a testing harness package (\texttt{imf\_test\_harness}, 4,251 LoC) exposing unified API interfaces for agents and interacts with both frameworks. 

The audit agent is a typical ReAct agent that dynamically invoke tools and decides next steps from its accumulated context. We adopted the code exploration tools from the implementation of Atlantis~\cite{kim:atlantis}.
On the other hand, the tester agent and the log analyzer agent both use fixed workflows, to make them compatible and can be directly invoked by the auditor, we also implemented them as LangGraph agents.

We use CodeQL to build the SliceDB with 7 \texttt{*.ql} files: 3 for \ardupilot and 4 for \px. The queries collect forward slices for MAVLink commands, messages, and uORB topics, and construct mappings between messages and commands for both frameworks, in total 1,846 LoC \texttt{*.ql} queries.

In the testing harness package, we implemented interfaces of 42 MAV\_CMDs, primarily focusing on mission navigation commands (e.g., \texttt{upload\_mission}, \texttt{upload\_mission\_waypoints}, \texttt{start\_mission}, etc.). 
Except for the interfaces we have already supported, we also implemented interfaces include \texttt{send\_param}, \texttt{send\_command}, and \texttt{send\_command\_int}, which allow agents to send arbitrary MAVLINK packets with directly providing the packet ID and parameters. This allows the agent to test some edge cases that unified interfaces cannot cover. 

Our implementation and findings are both released on Github anonymously. See the \S~\ref{sec:data-availability} Data Availability for details.

%% file: sections/06.evaluation.tex
\subsection{Overview}

We use GPT-5.4 for the auditor and the tester agents, and GPT-OSS-120B for the log analyzer. In the following sections, we first elaborate on end-to-end framework effectiveness (\S~\ref{sec:eval-e2e}), then discuss each components (\S~\ref{sec:eval-auditor}-\S~\ref{sec:eval-la}).

\subsection{Framework Effectiveness on Real-world Systems}
\label{sec:eval-e2e}

We evaluated \tool on \px and \ardupilot and uncovered 105 previously unknown logic flaws after
manual verification.
These flaws include complex cross-module state inconsistencies, improper input validation,
inconsistency between MAVLink protocol and implementatio, and other mission-logic failures, all of
which do not directly lead to crashes that can be captured by traditional sanitizers.

\vparstart{Analysis of Discovered Bugs.}
The audit agent proposed 428 candidate logic bug hypothesis findings. Before test generation, the
dependency checks filtered out 235 candidates that requires specific hardware backends to validate,
such as video capture pipelines, gimbal or mount controllers, winch gripper actuators. In SITL, these cases may expose command acceptance or state changes, but the decisive physical/backend-side effect cannot be observed.

The remaining 193 candidates proceeded to test-case generation and log analysis. 
The log analyzer filtered out 46 as false positives, and reported 147 as potential flaws. 
Two experts manually verified all the reported findings, inspected the hypotheses and execution logs,
and excluded 40 additional false positives. This left 107 confirmed logic-flaw reports. After merging two duplicate pairs, \tool identified 105 unique real logic flaws.
Table~\ref{tab:representative-bugs} in Appendix shows representative logic flaws, their
frameworks, MAVLink commands, and potential impact. The full list of the bugs \tool identified can
be found on our supporting
website\footnote{\href{https://ImpossibleMissionFinder.github.io}{https://ImpossibleMissionFinder.github.io}}.

\vparstart{Finding Categories.}
After manual validation, the confirmed logic flaws (Table~\ref{tab:finding_categories}) fall into
four categories: a) command input-handling defects, b) protocol incompliance, c) runtime state
inconsistencies, and d) safeguard enforcement failures.

\blt{Command input-handling defects}. 
57 bugs fall into this category. We found that many MAVLink command handlers do not properly
validate and handle command inputs. For instance, some handlers accept invalid values, such as
negative or NaN circle-fence radii, and store them as geofence parameters, which can make an
inclusion fence impossible to satisfy or weaken an exclusion fence. 
There are also handlers that accept parameters as valid but silently drop them during command
handling, leaving the GCS without a clear failure signal. We also found handlers that either miss
edge cases or contain numeric input-handling errors, where numeric parameters are converted or
truncated to other types before validation, causing precision loss or incorrect execution values.

\vblt{Protocol incompliance}. We found 11 cases where framework implementations do not follow
the MAVLink specification. For instance, the MAVLink protocol specifies \texttt{param4} of
\texttt{MAV\_\allowbreak CMD\_\allowbreak NAV\_\allowbreak LOITER\_\allowbreak TIME} as the
Xtrack Location field~\cite{mavlink-loiter-time}. 
However, \px also treats it differently and converts it into a yaw angle when parsing mission items. 
As a result, enabling xtrack exit with a positive \texttt{param4}
can unintentionally introduce a yaw requirement that changes loiter completion.

\vblt{Runtime state inconsistencies}. Similar to the example in \S~\ref{sec:motivation}, we found
29 cases where the runtime state variables are not synchronized across modules or execution paths.
For example, a race condition in \px pause/reposition handling can make the drone failure to pause
while still switches into the Hold/Loiter mode. The drone will keep flying toward the next waypoint
instead of stopping as the MAV\_CMD expected.

\vblt{Safeguard enforcement failures}. We found 8 logic bugs where the safeguards, such
as target component routing, geofence policy, or return-to-target validation are
not properly applied.
For example, some commands fail to validate GeoFence, allowing the drone to fly toward restricted
airspace.

\begin{table}[t]
\centering
\small
\caption{Categories of the logic flaws \tool identified.}
\label{tab:finding_categories}
\begin{tabular}{p{0.76\linewidth}r}
\toprule
\textbf{Category} & \textbf{Count} \\
\midrule
Command input-handling defects & 57 \\
Protocol incompliance & 11 \\
Runtime state inconsistencies & 29 \\
Safeguard enforcement failures & 8 \\
\midrule
\textbf{Total} & \textbf{105} \\
\bottomrule
\end{tabular}
\end{table}

\vparstart{False Positives.}
Of the 147 candidates reported by \tool, 40 were excluded as false positives after manual review.
Some hypotheses misinterpreted command or protocol semantics, for example by treating valid
special values as invalid inputs. Others overlooked upstream validation or downstream safeguards
that prevented the predicted failure. We also found hypotheses that combined execution paths whose
mode conditions could not hold simultaneously, making the proposed scenario infeasible. Finally,
normal transitions were sometimes mistaken for persistent failures, although subsequent entries in
the same execution log showed that the requested action completed under the allowed conditions or
within the permitted time.
We also found two duplicate pairs during manual review. The agent proposed similar hypotheses for
the same PX4 \texttt{MAV\_CMD\_CONDITION\_DISTANCE} execution gap and for the same PX4
geofence/return-target safety-invariant violation. We counted each pair once in the final total.

\subsection{Effectiveness of the Auditor}
\label{sec:eval-auditor}

To understand the effectiveness of the SliceDB in auditor agent, we perform an ablation study
comparing two different settings: a) \textit{Full} auditor agent with both indirect dataflow
analysis slicing and agent SliceDB tools enabled; and b) \textit{No SliceDB}, the same auditor agent
with SliceDB related tools removed, leaving only typical agent explorationg tools like
\texttt{read\_file}, \texttt{grep}, and \texttt{glob}, etc for code exploration. 
We prepared five gourps of MAVLink commands (Table~\ref{tab:auditor-ablation-inputs}) for the
auditor ablation evaluation, which are commands used in mission planning subsystems and have led to
existing bug issues. 
For the \textit{Full} setting, we run auditor and provide slicings of all the commands in the group
as part of the initial prompt. For the \textit{No SliceDB} setting, we provide the list of command
and directly prompt the model to find bugs related to the commands.
Table~\ref{tab:auditor-ablation-inputs}
lists the concrete command inputs and the existing issues that motivated each group.

\begin{table}[t]
\centering
\scriptsize
\caption{MAVLink command groups used in \S~\ref{sec:eval-auditor}}
\label{tab:auditor-ablation-inputs}
\begin{tabular}{@{}p{0.05\linewidth}p{0.6\linewidth}p{0.2\linewidth}@{}}
\toprule
\textbf{ID} & \textbf{MAVLink commands} & \textbf{Issues} \\
\midrule
G1 &
\texttt{DO\_SET\_HOME},\texttt{NAV\_LAND} 
\texttt{NAV\_RETURN\_TO\_LAUNCH}&
\#22373, \#22491\newline \#22576 \\
\addlinespace[0.25em]
G2 &
\texttt{DO\_FENCE\_ENABLE},
\texttt{DO\_FENCE\_DISABLE} &
\#22282, \#27099\newline ArduPilot \#25327 \\
\addlinespace[0.25em]
G3 &
\texttt{DO\_CHANGE\_ALTITUDE},
\texttt{DO\_REPOSITION}\newline
\texttt{NAV\_TAKEOFF},
\texttt{DO\_SET\_HOME} &
\#22491, \#22576 \\
\addlinespace[0.25em]
G4 &
\texttt{DO\_CHANGE\_ALTITUDE}\newline
\texttt{DO\_REPOSITION},
\texttt{DO\_CHANGE\_SPEED} &
\#22492\\
\addlinespace[0.25em]
G5 &
\texttt{NAV\_WAYPOINT},
\texttt{DO\_JUMP},
\texttt{NAV\_TAKEOFF} &
\#22255, \#27146 \\
\bottomrule
\end{tabular}
\end{table}

\vparstart{Result.}
On \px, the \textit{Full} audit agent produced 28 candidate findings while the \textit{No SliceDB}
agent produced 17.
On \ardupilot, the \textit{full} agent produced 23 candidates and the \textit{No SliceDB} agent
produced 22 findings.
The difference across frameworks reflects how each dispatches commands. 
\px's navigator routes a MAVLink command through several uORB publish/subscribe edges before the
consuming module sees it, and the pre-compute code slicing connects those indirect dataflows and
the slice-query tools guide the audit agent aware of the connections of those different code
snippets. Without the slicing tools, the agent reads one file at a time and tends to stop at
\texttt{orb\_publish} and cannot always locate the topic consumers.
\ardupilot, on the other hand, dispatches MAVLink commands through direct function calls, so
file-level reads are often enough for reconstructing the dataflow. 

\subsection{Effectiveness of the Tester}
\label{sec:eval-tester}
We evaluate whether the tester can automatically generate test cases for auditor-produced
hypotheses. We measure this ability with pass@$k$: a case is considered as pass at attempt $k$ if
the tester produces a valid test that has correct syntax, passes the semantic
check, and can be executed in the simulator within $k$ attempts,. 
We randomly sampled 50 hypotheses from the end-to-end evaluation results in
\S~\ref{sec:eval-e2e}. 
Table~\ref{tab:tester_passk} presents cumulative pass@$k$. 
The tester succeeds on 36/50 cases in the first attempt and reaches
46/50 by the third attempt.

\begin{table}[h]
\centering
\small
\caption{Tester pass@k on 50 randomly sampled candidate flaws.}
\label{tab:tester_passk}
\begin{tabular}{c|c|c}
\toprule
\textbf{$k$} & \textbf{Cumulative pass} & \textbf{Pass rate} \\
\midrule
1 & 36 / 50 & 72.0\% \\
2 & 43 / 50 & 86.0\% \\
3 & 46 / 50 & 92.0\% \\
4 & 46 / 50 & 92.0\% \\
\bottomrule
\end{tabular}
\end{table}

\vparstart{Failure analysis}.
The four remaining failures fall into two categories. Three cases were rejected by the semantic
sanity check because the generated tests do not properly meet the requirements. For example, some
hypotheses require constructing malformed MAVLink messages while the tester used the unified API
interfaces instead of the API that can send raw messages (\texttt{send\_command}).
The remaining failure was an infrastructure issue: the generated PX4 arm/disarm test passed
semantic checking, but the execution connection timed out before producing a parseable log.

\subsection{Effectiveness of the Log Analyzer}
\label{sec:eval-la}

To understand the effectiveness of the Log Analyzer, we compare three different settings on a
curated dataset: log analyzer agent, log analyzer agent without auditor hypothesis, and a
log-anomaly baseline (Dronlomaly~\cite{shar2022dronlomaly}) work.

\vparstart{Setup.}
We construct the dataset with 50 logs, balanced 25 anomaly + 25 benign and 25 \px + 25 \ardupilot. 
The 25 anomaly logs are replayed from candidates whose flaws were confirmed in
\S~\ref{sec:eval-e2e}, and the 25 benign logs come from execution log of frameworks' public unit test
execution logs and from logs of \tool confirmed as false positives in \S~\ref{sec:eval-e2e}.
Dronlomaly is compared as the baseline method: it trains a machine-learning classifier over selected
features, such as the \texttt{x/y/z} velocity and \texttt{x/y/z} acceleration. We adopt the original
setup in Dronlomaly including their model architecture, parameters, and default threashold following
the original paper. 
The log analyzer without hypothesis setting keeps the core ego-state and the sliding window
unchanged, but it does not know the logic flaw hypothesis auditor proposed.

\begin{table}[h]
\centering
\small
\caption{Log analyzer detection results on the 50-case dataset (25 anomaly + 25 benign).}
\label{tab:la_results}
\begin{tabular}{l|c|c|c|c}
\toprule
\textbf{Configuration} & \textbf{TP} & \textbf{FP} & \textbf{FN} & \textbf{F1} \\
\midrule
Log analyzer & 19 & 3 & 6 & 0.81 \\
Log Analyzer without Hypothesis & 18 & 11 & 7 & 0.67 \\
Dronlomaly & 4 & 15 & 21 & 0.18 \\
\bottomrule
\end{tabular}
\end{table}

\vparstart{Results.}
Table~\ref{tab:la_results} shows the results across the three configurations.
The full log analyzer achieves the highest F1 (0.81) because it uses the auditor's hypothesis as a
candidate-specific oracle and checks whether the predicted behavior appears in the runtime log.
Without this hypothesis, the log analyzer falls back to a generic anomaly judgment and produces
more false positives. 
Dronlomaly is even less aligned with our flaws because its classifier models only checks low-level
motion features and can not understand high-level semantic inconsistency at all, making it not
possible to detect most of the logic bugs.

\subsection{Comparison with Existing Frameworks}
\label{sec:eval-compare-with-pgfuzz-adgfuzz}

There are several prior works target logic-vulnerability discovery in autopilot 
software~\cite{kim2021pgfuzz,kim2019rvfuzzer,wangADGFUZZAssignmentDependencyGuided}, but none of
them can uncover the types of logic flaws \tool identified. 
PGFuzz~\cite{kim2021pgfuzz} can find policy-guided violations, but
it requires human experts to encode the expected behavior as temporal-logic policies. Writing these
policies depends on detailed knowledge of both the code and the intended vehicle semantics, and is
both time-consuming and error-prone. 
ADGFuzz~\cite{wangADGFUZZAssignmentDependencyGuided} focuses on assignment-dependency bugs, i.e.,
faults whose root cause can be exposed around variable assignments, and cannot cover a large portion
of logic flaws we found, such as protocol-compliance issues and complex internal-state inconsistencies.

\tool's advantage comes from using LLMs to identify candidate logic flaws hypotheses, making it more
flexible and not limit to fixed bug patterns. 
The tester enables automatic validation and executes test cases dynamically.
The log analyzer then checks whether the logic flaws hypothesis
actually appears in runtime logs, giving \tool an automated validation mechanism for diverse
semantic bugs rather than a manually written oracle for one bug class.

%% file: sections/07.case_study.tex
\begin{figure}[b]
\centering
\includegraphics[width=.95\textwidth]{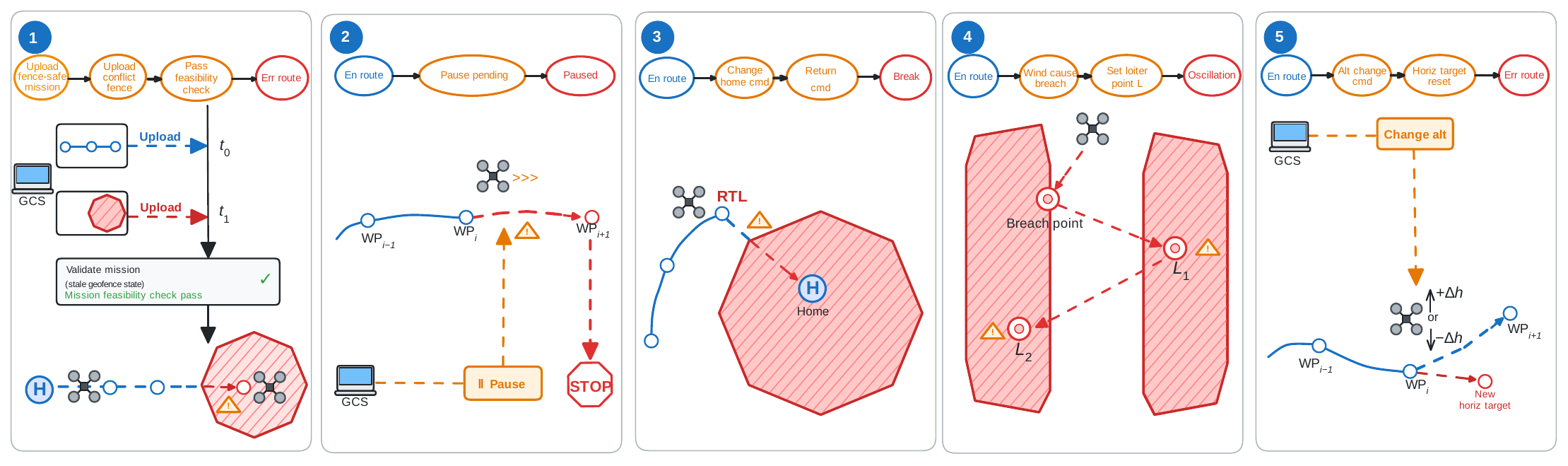}
\caption{Overview illustration of the five representative case studies discussed in
         \S~\ref{sec:case_study}:~ {\color{black}{\one}} No-Fly Zone not properly enforced;
         {\color{black}{\btwo}} Drone cannot be properly paused; {\color{black}{\bthree}} Drone can return to
         locations in No-Fly Zone; {\color{black}{\four}} Drone trapped inside narrow no fly zones; {\color{black}{\five}} Changes of altitude of home position leading to incorrect waypoint reachability check;}
\label{fig:case-study-overview}
\end{figure}

To illustrate the breadth of logic flaws found by \tool, this section presents five representative
cases from \S~\ref{sec:evaluation}. 
Figure~\ref{fig:case-study-overview} provides a compact visual summary of their scenario structure.
These cases do not follow a single pattern: they span stale validation, race conditions,
inconsistent state updates, flawed recovery logic, and command-semantics errors. 
Together, they demonstrate the effectiveness and practicality of \tool in discovering real-world
complex logic flaws in autopilot mission planning and handling modules.

\vspace{3pt}\noindent \textbf{Case {\color{black}{\one}}: Mission not rechecked after geofence update.}
\px checks an uploaded mission against the current geofence configuration before allowing the
mission to run. The flaw appears when the operator first uploads a valid mission and later uploads a
new no-fly zone that covers part of that mission path. The new fence is accepted, but the already
loaded mission is not reliably rechecked against the updated fence state. Consequently, the autopilot
continues to execute a mission that was valid under the old safety policy but invalid under the new
one, and the vehicle can fly into the newly defined restricted region.
An operator may upload a new fence to keep the drone away from a temporary hazard.
In this case, the upload succeeds, but the drone can still enter that area while following
its previously loaded mission.

\vparstart{Case {\color{black}{\btwo}}: Pause command race condition.}
In mission mode, a pause or hold command should replace the active mission target with a stop/hold
target near the current vehicle position. \tool found a race condition in this mechanism: after the
operator issues the pause command, one part of the system records the new pause state, while the
navigator can still use the previous mission waypoint target. In the tested scenario, the vehicle
therefore acknowledged the pause-related state change but continued flying toward the next waypoint
instead of stopping. The bug is caused by inconsistent internal state across command handling and
mission navigation, not by a missing command handler.
If the operator pauses a mission to avoid an obstacle, the drone may keep approaching it
even though its mode has changed to Hold/Loiter. The mode change alone would give the operator
no indication that the pause had failed.

\vparstart{Case {\color{black}{\bthree}}: RTL to a home point inside a no-fly zone.}
\texttt{MAV\_CMD\_DO\_SET\_HOME} updates the vehicle's home position, and RTL uses that home position
as its recovery destination. \tool found that \px does not consistently validate the updated home
position against active no-fly zones before RTL uses it. In the scenario, the vehicle is already in
flight, the operator changes the home position to a location inside a no-fly zone, and the vehicle
later switches to RTL. Because RTL trusts the stored home position as a safe target, the recovery
mode can guide the drone directly into restricted airspace.
The effect of the home update may only become apparent when RTL starts later in the flight.
An operator relying on RTL to bring the drone back could therefore send it into a no-fly zone.

\vparstart{Case {\color{black}{\four}}: Oscillation between two no-fly zones.}
\px geofence recovery tries to push the vehicle back out of a no-fly zone after a boundary breach.
This local recovery rule becomes unsafe in a narrow corridor between two no-fly zones. If wind,
inertia, or the waypoint route pushes the vehicle into the front restricted region, the recovery
logic commands a backward escape motion. That motion can carry the vehicle into the no-fly zone
behind it, which triggers another recovery command in the opposite direction. The two locally
reasonable recovery actions can therefore interact and keep the drone moving back and forth between
the two restricted regions.
The drone can remain trapped between the two zones rather than resume its mission.
The operator may need to take over to move it out of the repeated recovery maneuvers.

\vparstart{Case {\color{black}{\five}}: Home-altitude update breaks waypoint reachability.}
\texttt{MAV\_CMD\_DO\_SET\_HOME} can update not only the home location but also the home altitude, and
\px uses home-relative altitude when evaluating mission waypoint reachability. The flaw appears when
the home altitude is changed while a mission is already running. After the update, part of the
navigator still follows mission state computed from the previous home altitude, while the
reachability check evaluates the current waypoint using the updated home altitude. The vehicle can
therefore be near the intended waypoint in physical space but fail the internal reachability test,
which can stall waypoint completion or otherwise disrupt mission progress.
The drone may remain airborne near the waypoint without advancing to the next mission item.
Later parts of the mission are then blocked because the autopilot still considers the current
waypoint incomplete.

\vparstart{Additional representative cases.}
There are also representative protocol incompliance cases and input handling cases. 
For instance, \ardupilot accepts a \texttt{MAV\_CMD\_NAV\_PAYLOAD\_PLACE} mission item at
latitude/longitude $(0,0)$ as a valid stored location, but the execution path later treats the same
coordinate as ``no location provided'' and starts payload placement at the vehicle's current
position. 
Another example is \texttt{MAV\_CMD\_DEBUG\_TRAP}: the handler ignores \texttt{target\_component},
so a debug-only command addressed to another component can still trigger the trap on the flight
controller. 
A third example, \px mission \texttt{MAV\_CMD\_COMPONENT\_ARM\_DISARM}, shows a
parameter-propagation failure: the mission path accepts the command but drops the force flag, so the
later executor sees a different disarm request than the GCS sent. These cases complement the five
figure cases because they cover protocol interpretation, authority checks, and command-parameter
propagation rather than another variant of geofence recovery.

%% file: sections/08.discussion.tex
\vparstart{Hardware Dependency.}
\tool includes a testing package that provides unified interfaces for agents to write test cases
and execute them automatically inside simulators. Admittedly, this design cannot find logic bugs that
depend on specific hardware backends (e.g., cameras, actuators, or gimbals). \tool is designed to
identify logic flaws in general modules shared across drones, such as mission control and command
handling logic, which are central parts of autopilot software and can be tested in simulation.

\vparstart{Generalizability.}
While we focus on finding logic bugs within autopilot frameworks, the high-level design of the
pipeline can transfer to other systems:
preparing curated code slices to guide LLMs toward targeted code context, generating and executing
LLM-written test cases for runtime validation, and analyzing logs with an ego-state representation
are all broadly applicable.
\tool currently targets bugs caused by MAVLink command handling and mission-related logic;
lower-level controllers, sensor drivers, GPS spoofing, and other physical-layer attacks are out of
scope, and automatically identifying logic bugs in those modules is left as future work.

\vparstart{LLM dependence.}
\tool is built as a multi-agent system, and all three modules rely on the performance of LLMs.
As models become more powerful, we expect the audit agent to reason over longer and more complex
code slices, the tester and log analyzer agent to be more accurate.
At the same time, this dependence means that \tool's effectiveness may vary across models and
reaoning efforts. 
We mitigate this risk by executing test cases in simulator and validating the execution logs. LLMs
are used to guide hypothesis generation and interpretation, but the final evidence still comes from
executable tests and observed system behavior. We also manually validated all the reported flaws
\tool identified and only reported those confirmed by human experts.

%% file: sections/09.related_works.tex
\noindent\textbf{Bug finding for cyber-physical and embedded software.}
Classical static analysis, symbolic execution, and fuzzing~\cite{chess2004static,cadar2008klee,klees2018evaluating}, as well as modular dependence analysis for large C/C++ systems~\cite{lu2023practical}, are effective for localized memory-safety, protocol, or invariant violations. \tool targets a different signal: mission-logic flaws whose evidence crosses runtime-resolved publish-subscribe or dispatch edges and whose symptom is semantic divergence rather than a crash.

\smallskip\noindent\textbf{Autopilot fuzzing.}
PGFuzz~\cite{kim2021pgfuzz}, PGPatch~\cite{kim2022pgpatch}, PatchVerif~\cite{kim2023patchverif}, LGDFuzzer~\cite{lgdfuzzer}, ADGFUZZ~\cite{wangADGFUZZAssignmentDependencyGuided}, and RVFuzzer~\cite{kim2019rvfuzzer} tailor testing to drone or robotic-vehicle software. These systems rely on hand-written policies, differential patch checks, assignment dependencies, environmental mutations, or other predefined oracle spaces. \tool instead asks the auditor to synthesize a candidate-specific behavioral divergence from slice context, then validates that divergence in SITL.

\smallskip\noindent\textbf{Runtime and log-based anomaly detection.}
Drone log forensics and anomaly detectors~\cite{shar2022dronlomaly,baig2022drone-forensics,kumar2021drone-forensics,silalahi2023transformer-forensics} reconstruct flights or detect deviations from learned normal behavior, and recent LLM-based log-anomaly work extends that paradigm~\cite{he2024llmelog,almodovar2024logfit,pan2024raglog,karlsen2024benchmarking}. \tool's log analyzer is not a generic anomaly detector: it consumes the candidate's predicted divergence and checks whether that exact relation appears in the run, which reduces false positives from benign trajectory variation.

\smallskip\noindent\textbf{LLM agents for code analysis and bug finding.}
LLM-for-code work~\cite{brown2020language,chen2021evaluating} and follow-on systems for vulnerability detection, test generation, unit testing, and smart-contract auditing~\cite{llmbugdetect1,llmbugdetect2,schafer2023empirical,wei2024llm,li2025everything} show that LLMs can reason about code but often struggle with noisy whole-file context and unverified findings. \tool adds two domain-specific safeguards: slice context that crosses autopilot indirection and dynamic confirmation that turns static hypotheses into executable evidence.

%% file: sections/11.appendix.tex
\begingroup
\footnotesize
\setlength{\tabcolsep}{3pt}
\newlength{\imfCommandWidth}
\settowidth{\imfCommandWidth}{\texttt{MAV\_CMD\_NAV\_FENCE\_CIRCLE\_INCLUSION}}
\begin{longtable}{@{}p{12pt}p{34pt}lp{\dimexpr\linewidth-46pt-\imfCommandWidth-6\tabcolsep\relax}@{}}
\caption{Representative logic flaws identified by \tool. The full list is available on our \href{https://ImpossibleMissionFinder.github.io}{supporting website}.}
\label{tab:representative-bugs}\\
        \toprule
        \textbf{ID} & \textbf{Platform} & \textbf{Related MAVLink Command} & \textbf{Description} \\
        \midrule
        \endfirsthead
        \toprule
        \textbf{ID} & \textbf{Platform} & \textbf{Related MAVLink Command} & \textbf{Description} \\
        \midrule
        \endhead
        \bottomrule
        \endfoot
        1 & \px & \texttt{MAV\_CMD\_COMPONENT\_ARM\_DISARM} &
        A mission can request a forced disarm by sending a magic force value with the arm/disarm command. PX4 accepts such a mission item but drops the force parameter before execution, so Commander later evaluates it as an ordinary disarm request. \\[0.15ex]
        2 & \ardupilot & \texttt{MAV\_CMD\_CONDITION\_DISTANCE} &
        This command is meant to delay following mission actions until the vehicle is close enough to the next waypoint. ArduPilot accepts negative or fractional distances, then stores the threshold in integer runtime state, which can make the condition pass at the wrong time. \\[0.15ex]
        3 & \px & \texttt{MAV\_CMD\_NAV\_LOITER\_TIME} &
        A loiter-time mission item should make the vehicle wait for a time interval, with a separate field controlling cross-track exit behavior. PX4 also interprets that field as a yaw target, so enabling the exit flag can accidentally add a yaw-completion requirement. \\[0.15ex]
        4 & \ardupilot & \texttt{MAV\_CMD\_NAV\_PAYLOAD\_PLACE} &
        A mission can ask the vehicle to fly to a location and place a payload there. ArduPilot stores latitude/longitude $(0,0)$ as a valid target during upload, but the executor later treats the same coordinate as a sentinel meaning no target was provided. \\[0.15ex]
        5 & \px & \texttt{MAV\_CMD\_DO\_REPOSITION} &
        Ground stations use this command to pause or reposition a vehicle during a mission. A race between PX4's commander and navigator can switch the visible mode to Hold/Loiter while the active navigation target still points to the next mission waypoint. \\[0.15ex]
        6 & \ardupilot & \texttt{MAV\_CMD\_DEBUG\_TRAP} &
        This debug-only command can intentionally trap the autopilot process when sent with a magic value. ArduPilot's handler does not check the target component, so a command addressed to another onboard component can still affect the flight controller. \\[0.15ex]
        7 & \px & \texttt{MAV\_CMD\_DO\_SET\_HOME} &
        Home is the reference point for relative-altitude missions. PX4 allows Home to change during flight, so an already-active waypoint can be reinterpreted against the new altitude reference and may no longer satisfy its completion check. \\[0.15ex]
        8 & \px & \texttt{MAV\_CMD\_NAV\_FENCE\_CIRCLE\_INCLUSION} &
        A circular inclusion fence is intended to define an allowed flight area with a positive radius. PX4 accepts zero or negative radii, which can make the allowed region impossible to satisfy or make later fence checks inconsistent with the uploaded policy. \\[0.15ex]
        9 & \ardupilot & \texttt{MAV\_CMD\_CONDITION\_DELAY} &
        This command delays subsequent mission actions for a requested number of seconds. ArduPilot accepts negative delay values during mission upload and later executes them as an immediate pass-through, so the next action can run without the intended wait. \\[0.15ex]
        10 & \ardupilot & \texttt{MAV\_CMD\_DO\_REPEAT\_SERVO} &
        The command repeats a servo movement a requested number of times. ArduPilot accepts negative repeat counts and can reinterpret them as very large counts, turning a bounded servo action into an effectively unbounded oscillation. \\[0.15ex]
\end{longtable}
\endgroup

%% file: sections/conclusion.tex
Leveraging LLMs for logic flaw discovery in autopilot frameworks requires high-quality context
for LLMs, runtime verification infrastructure to confirm candidate flaws, and semantic bug oracles
for detecting different kinds of logic violations.
In this paper, we present \tool, a multi-agent system that automatically
\underline{\textit{identifies}}, \underline{\textit{tests}}, and \underline{\textit{validates}} logic
flaws in autopilot software.
\tool addresses these challenges by constructing command-level code slices as targeted LLM context,
turning LLM-generated hypotheses into simulator tests through a unified testing interface, and
validating each execution with an LLM-based log analyzer that checks whether the candidate-specific
runtime violation appears in the execution logs.
We comprehensively evaluated \tool on \px and \ardupilot, uncovering 105 previously unknown
logic flaws after deduplication and demonstrating the effectiveness of the framework.